# Two Loop Computation of a Running Coupling in Lattice Yang-Mills Theory

Rajamani Narayanan
School of Natural Sciences, Institute for Advanced Study
Olden Lane, Princeton, NJ 08540, USA

Ulli Wolff
Institut für Physik, Humboldt Universität
Invalidenstr. 110, D-10099 Berlin, Germany

**Abstract**
We compute the two loop coefficient in the relation between the lattice bare coupling and the running coupling defined through the Schrödinger functional for the case of pure SU(2) gauge theory. This result is needed as one computational component to relate the latter to the $\overline{\text{MS}}$-coupling, and it allows us to implement $O(a)$ improvement of the Schrödinger functional to two-loop order. In addition, the two-loop $\beta$-function is verified in a perturbative computation on the lattice, and the behavior of an improved bare coupling is investigated beyond one loop.

# 1 Introduction

The lattice formulation of pure gauge theories and QCD enables one, in principle, to connect the non-perturbative (low energy) regime to the perturbative (high energy) regime. An important example is the computation of the running coupling at short distances in some scheme, say minimal subtraction (MS), with the scale set by a low energy quantity of the theory. At the outset there are two problems that make this computation practically difficult. One is the apparent need for large lattices to span a wide range of length scales. This problem can be circumvented by the use of finite-size scaling as was first demonstrated for non-linear sigma models in two-dimensions [1]. The second problem is to find a physical renormalized quantity to serve as a running coupling, that has a number of favorable properties [2]. Firstly, it should allow for an accurate numerical estimation on a range of lattices. Secondly, it should exhibit small cut-off dependence, so that a reliable and precise extrapolation to the universal continuum values is possible from medium size lattices. Thirdly, although the definition has to be non-perturbative, the quantity must have a manageable perturbation expansion at high energy. This is required to perturbatively relate the coupling chosen for the simulation to other more conventional ones at high energy.

In refs. [3], [4], [5] the coupling $\bar{g}^2(L)$, based on the Schrödinger functional with an induced abelian background field, was found to fulfill these requirements to a sufficient degree. In the present paper, for the case of SU(2), we work out the 2-loop contribution to $\bar{g}^2$ on the lattice in terms of the bare coupling $g_0^2 = 4/\beta$. Recently [6] a calculation of the relation between $g_0^2$ and $g_{\overline{\mathrm{MS}}}^2$ was reported and combined with our result. We have thus succeeded in connecting $\bar{g}^2$ and $g_{\overline{\mathrm{MS}}}^2$ at our highest energy with an accuracy needed to match the improved statistical precision on parallel computers [7]. In addition, the present computation alone allows us to implement O($a$) Symanzik improvement of our action and coupling to 2-loop order. As a by-product, we investigate the expansion of $\bar{g}^2$ in terms of a simple plaquette-improved bare coupling [8] to one more order than before. This is highly welcome information, as up to now, only the smallness of 1-loop contributions in terms of such couplings could be verified for some quantities.

The organization of the paper is as follows. In sect. 2 we briefly remind the reader of the definition of $\bar{g}^2$ in terms of the Schrödinger functional and set up the necessary expansions. In sect. 3 we list perturbative results for our coupling on a series of lattices and analyze their cut-off dependence. Up to here the paper should be readable in a self-contained fashion for a reader, who just wants to understand the problem and the result. The remainder contains more technical material of interest to those who plan a similar computation. In that part we assume that the reader has a copy of [3] at hand, and we refer to individual equations and sections of this reference. In sections 4 and 5 we discuss the structure of the gauge fixed action and the construction of propagators in the background field. In sect. 6 all terms required at 2-loop order are collected and the optimization of the computational speed is discussed.

## 2 Running coupling and Schrödinger functional

In this section we summarize the steps in defining the Schrödinger functional based running coupling $\bar{g}^2$ proposed in [3] and define its perturbative expansion. We use lattice units, setting the spacing $a$ to unity, and choose a box with temporal and spatial extent $L$.

The Schrödinger functional is the kernel for euclidean propagation from some field configuration at time $x^0 = 0$ to some other configuration at $x^0 = L$. In the path integral formalism, this becomes the partition function on an $L^4$ box with fixed boundary conditions at $x^0 = 0, L$ and periodic ones in the three space directions.

A lattice gauge field $U$ is an assignment of a link variable $U(x, \mu) \in \mathrm{SU}(2)$ to every pair $(x, x + \hat{\mu})$ of lattice points. In the path integral to be defined, all gauge fields in the interior fluctuate freely, and the ones on the spatial links in the time slices $x^0 = 0, L$ remain fixed at

$$U(x,k)|_{x^0=0} = \exp\{C_k\}; \quad U(x,k)|_{x^0=L} = \exp\{C'_k\}; \quad k = 1,2,3, \qquad (2.1)$$

with

$$C_k = \frac{\eta}{iL}\tau^3; \quad C'_k = \frac{(\pi - \eta)}{iL}\tau^3; \quad \tau^3 = \begin{pmatrix} 1 & 0 \\ 0 & -1 \end{pmatrix}. \qquad (2.2)$$

These boundary conditions contain one dimensionless free parameter with range [3] $0 < \eta < \pi$. The Schrödinger functional is now given by

$$\mathcal{Z}(\eta, L) = \int D[U] e^{-S(U)}, \quad D[U] = \prod_{x,\mu} dU(x,\mu), \qquad (2.3)$$

where one integrates over all interior lattice gauge fields with the fixed boundary values given by eq. (2.1). The action $S[U]$ is taken to be

$$S[U] = \frac{1}{g_0^2} \sum_{x,\mu,\nu} w_{\mu\nu}(x) \text{tr}\{1 - U_{\mu\nu}(x)\}, \qquad (2.4)$$

where the sum runs over all oriented plaquettes on the lattice and $U_{\mu\nu}(x)$ denotes the parallel transporter around the plaquette. The weight $w_{\mu\nu}(x)$ is given by

$$w_{\mu\nu}(x) = \begin{cases} \frac{1}{2} c_s(g_0) & \text{for spatial plaquettes at } x^0 = 0 \text{ or } x^0 = L, \\ c_t(g_0) & \text{for time-like plaquettes touching the boundary,} \\ 1 & \text{elsewhere.} \end{cases} \qquad (2.5)$$

In the classical limit the choice $c_s(0) = c_t(0) = 1$ eliminates $O(a)$ lattice artifacts. Beyond it, the functions $c_s(g_0)$ and $c_t(g_0)$ are chosen to minimize cut-off effects. For the abelian boundary fields (2.2), $c_s(g_0)$ is irrelevant and will hence not be discussed further (see however [3] for 1-loop values of both weights). For perturbative Symanzik improvement we insert

$$c_t(g_0) = 1 + c_t^{(1)} g_0^2 + c_t^{(2)} g_0^4 + \cdots \qquad (2.6)$$

and will report the 2-loop coeffient $c_t^{(2)}$ in the next section.

With the present boundary conditions the gauge field

$$V(x,0) = 1; \quad V(x,k) = \exp\left\{\left[x^0 C_k' + (L - x^0) C_k\right]/L\right\} \qquad (2.7)$$

and its gauge transforms minimize the classical action with a value

$$\Gamma_0(\eta, L) = g_0^2 S[V]\Big|_{g_0^2 = 0} = 24 L^4 \sin^2\left\{\frac{\pi - 2\eta}{2L^2}\right\}. \qquad (2.8)$$

The response to an infinitesimal change in $\eta$ defines the renormalized coupling proposed in [3],

$$\bar{g}^2(L) = \frac{\Gamma_0'(\eta, L)}{\Gamma'(\eta, L)}\bigg|_{\eta = \pi/4}, \qquad (2.9)$$

where

$$\Gamma(\eta, L) = -\ln \mathcal{Z}(\eta, L) \qquad (2.10)$$

is the effective action. The primes denote derivatives with respect to $\eta$. Here, and in all simulations sofar, the value $\eta = \pi/4$ has been chosen [3], which gives a good signal in the Monte Carlo simulation.

In perturbation theory in $g_0$ the expansion of the effective action

$$\Gamma = g_0^{-2}\Gamma_0 + \Gamma_1 + g_0^2\Gamma_2 + \cdots \tag{2.11}$$

yields

$$\bar{g}^2(L) = g_0^2 + m_1(L)g_0^4 + m_2(L)g_0^6 + \cdots, \tag{2.12}$$

where

$$m_1 = -\frac{\Gamma'_1}{\Gamma'_0}; \quad m_2 = -\frac{\Gamma'_2}{\Gamma'_0} + m_1^2. \tag{2.13}$$

The computation of $m_2(L)$ is the subject of this article.

## 3 Results

The perturbative expansion of the Schrödinger functional on the lattice with its finite induced background field can only be evaluated numerically. We carry this out for a range of lattices up to $L = 32$ and analyze the asymptotic $L$-dependence of the various expansion coefficients of interest. The general technique follows closely what is described in [10], and details of our calculation are given in the next sections. Here we only mention that the numbers we report here underwent a battery of tests like independence of a free gauge fixing parameter and verification of symmetries before actually using them to shorten the sums. Moreover, the two authors created two completely independent programs (separated by the atlantic ocean), which yielded identical results up to $L = 8$. Beyond this size, we only ran the program that has been optimized for speed. The total effort is on the order of $\sim 1000$ h on a Sun 10-41.

### 3.1 Coefficients for the coupling from $L = 4\ldots32$

To exhibit the dependence on improvement we parameterize $m_1(L), m_2(L)$,

$$m_1 = m_1^a + c_t^{(1)}m_1^b \tag{3.1}$$

$$m_2 = m_2^a + c_t^{(1)}m_2^b + \left[c_t^{(1)}\right]^2 m_2^c + c_t^{(2)}m_2^d + m_1^2 \tag{3.2}$$

where the occurring powers of $c_t^{(i)}$ follow from a glance at the structure of the perturbative terms (sect. 6). Some of the contributions are simple and can be given in closed form

$$m_1^b = m_2^d = -\frac{2}{L} \tag{3.3}$$

$$m_2^c = \frac{L-2}{L^2}\left[1 + \cos^{-2}\left(\frac{\pi - 2\eta}{L^2}\right)\right]. \tag{3.4}$$

Numerical results for the remaining terms are listed in table 1. Values for $m_1(L)$ were partly quoted[1] in [3] and are extended here for completeness. They are known in extended precision (128 bits), and all digits in the table are significant. The 2-loop computation was done in double precision (64 bits), and the truncation is such that the expected error is in the last digit.

## 3.2 Perturbation theory of the step scaling function

In our previous work [1] – [5] we introduced the step scaling function to characterize the evolution of a renormalized coupling under a finite rescaling factor $s$. For $\bar{g}^2$ it is defined in the continuum by

$$\sigma(s, u) = \bar{g}^2(sL)\,|_{u=\bar{g}^2(L)}. \tag{3.5}$$

On the lattice, this universal function [3] emerges in the limit from pairs of lattices of sizes $L, sL$ which diverge in lattice units,

$$\sigma(s, u) = \lim_{1/L \to 0} \Sigma(s, u, 1/L). \tag{3.6}$$

Both $\Sigma$ and $\sigma$ possess a perturbative expansion in $u$. Of particular interest for the numerical work is the perturbative expansion of the effect of the lattice,

$$\delta(u, 1/L) = \frac{\Sigma(2, u, 1/L) - \sigma(2, u)}{\sigma(2, u)} = \delta_1(u, 1/L)\,u + \delta_2(u, 1/L)\,u^2 + \ldots. \tag{3.7}$$

The coefficients $\delta_1, \delta_2$ can be straightforwardly derived from $m_1, m_2$ and are collected in table 2. We notice that the 1-loop improved $\delta_1^{(1)}$ and the 2-loop improved $\delta_2^{(2)}$ decay roughly proportional to $1/L^2$ as expected.

---

[1] In [3], $m_1(L)$ was defined not to include the contribution from 1-loop improvement and corresponds to $m_1^a(L)$ here.

| $L$ | $m_1^a$ | $m_2^a$ | $m_2^b$ |
|---|---|---|---|
| 4  | .3130122116558318 | .0184943911496276 | .0664683049711990 |
| 5  | .3356559317382983 | .0201529604832220 | .0630548432288889 |
| 6  | .3542203018055683 | .0214004662236067 | .0594592734230283 |
| 7  | .3700176237183195 | .0223562763764735 | .0560224083279087 |
| 8  | .3837042882617212 | .023114131327289  | .052832729952602  |
| 9  | .3957339261476999 | .023736097645165  | .049919595125402  |
| 10 | .4064416740603939 | .024261251647286  | .047279532208960  |
| 11 | .4160777750705229 | .024714638201379  | .044892902842661  |
| 12 | .4248310669113876 | .025112969750626  | .042734265949594  |
| 13 | .4328461895403340 | .025467870113241  | .040777668189208  |
| 14 | .4402357794126316 | .025787718323464  | .038998959271153  |
| 15 | .4470889876397987 | .026078735979548  | .037376602004327  |
| 16 | .4534774495677516 | .02634565788463   | .035891783962232  |
| 17 | .4594595194472748 | .02659216313431   | .034528233650916  |
| 18 | .4650833216291134 | .02682116212734   | .033271932732646  |
| 19 | .4703889852141854 | .02703499336957   | .032110811438287  |
| 20 | .4754103068101125 | .02723556184632   | .031034464031614  |
| 21 | .4801760063068773 | .02742443844283   | .030033897483170  |
| 22 | .4847106884542001 | .02760293275129   | .029101315688328  |
| 23 | .4890355885758341 | .02777214729876   | .028229936942690  |
| 24 | .4931691576726372 | .02793301855603   | .027413840712453  |
| 25 | .4971275264639118 | .02808634837864   | .026647839420229  |
| 26 | .5009248770736207 | .02823282841524   | .025927371235302  |
| 27 | .5045737434715104 | .02837305927392   | .025248410342669  |
| 28 | .5080852563831408 | .02850756573057   | .024607391692524  |
| 29 | .5114693444997410 | .02863680891327   | .024001147727234  |
| 30 | .5147349009901648 | .02876119615134   | .023426855017935  |
| 31 | .5178899222329467 | .02888108900318   | .022881989110948  |
| 32 | .5209416241340969 | .0289968098474    | .022364286191250  |

Table 1: List of $L$-dependent coefficients $m_1^a, m_2^a$ and $m_2^b$.

| $L$ | $\delta_1^{(0)}$ | $\delta_1^{(1)}$ | $\delta_2^{(0)}$ | $\delta_2^{(1)}$ | $\delta_2^{(2)}$ |
|---|---|---|---|---|---|
| 4 | 0.00631 | -0.00726 | 0.00255 | 0.00224 | -0.00064 |
| 5 | 0.00641 | -0.00445 | 0.00204 | 0.00194 | -0.00036 |
| 6 | 0.00623 | -0.00282 | 0.00163 | 0.00168 | -0.00024 |
| 7 | 0.00584 | -0.00192 | 0.00132 | 0.00147 | -0.00017 |
| 8 | 0.00540 | -0.00139 | 0.00109 | 0.00131 | -0.00012 |
| 9 | 0.00497 | -0.00106 | 0.00091 | 0.00118 | -0.00009 |
| 10 | 0.00459 | -0.00084 | 0.00077 | 0.00108 | -0.00007 |
| 11 | 0.00425 | -0.00068 | 0.00066 | 0.00099 | -0.00006 |
| 12 | 0.00396 | -0.00056 | 0.00057 | 0.00091 | -0.00005 |
| 13 | 0.00370 | -0.00048 | 0.00050 | 0.00085 | -0.00004 |
| 14 | 0.00347 | -0.00041 | 0.00044 | 0.00079 | -0.00003 |
| 15 | 0.00327 | -0.00035 | 0.00038 | 0.00074 | -0.00003 |
| 16 | 0.00309 | -0.00031 | 0.00034 | 0.00070 | -0.00002 |

Table 2: Perturbative lattice effects in the step scaling function, eq. (3.7). The superscripts in brackets here refer to 0, 1 and 2-loop $O(a)$ improvement.

## 3.3 Asymptotic analysis

In the spirit of ref. [10] we now derive the asymptotic expansion of $m_1$ and $m_2$ in terms of logarithmically modified powers of the lattice spacing, $L^{-m}(\ln L)^n$, with $n \leq 2$ for our 2-loop calculation. As stressed in [10], the generally best way to access the successive terms in the expansion is to recursively take discrete $L$-differences in such a way, that the desired term survives, while all the more leading ones and some of the next subleading ones are canceled. Over a range of $L$ the approximate constancy (up to uncancelled subleading terms) of this combination has to be checked. In this process both the available range of $L$ as well as the number of significant digits shrink, which limits the degree of cancellation possible. An analysis of both systematic [10] and roundoff errors has been implemented. It also has to be born in mind, that the expansion in $1/L$ is expected to be asymptotic only. We saw indications that for $L \leq 32$ the degree in $1/L$ that we could access with our 64-bit precision, roughly matches with the number of terms where the expansion makes sense. This means, that to reduce the errors in the numbers to follow, we would need both more digits and larger $L$. Neither of them alone would help significantly. An extension in this direction — fortunately not urgently needed — would seem rather difficult, as there seem to be hardly any compute-resources available that are efficient for large scale number crunching in 128-bit precision.

The result for $m_1$ in [3] is

$$m_1^a(L) = 2b_0 \ln L + 0.202349(3) - 0.1084(11)\frac{1}{L} + \mathrm{O}\left(\frac{\ln L}{L^2}\right), \qquad (3.8)$$

where the first of the universal coefficients of the $\beta$-function

$$b_0 = \frac{11}{24\pi^2}, \quad b_1 = \frac{17}{96\pi^4}, \qquad (3.9)$$

enters. Demanding the absence of a $1/L$ artifact in $m_1$ fixes the 1-loop improvement coefficient [3]

$$c_t^{(1)} = -0.0543(5). \qquad (3.10)$$

For $m_2^a$ we find

$$m_2^a(L) = 2b_1 \ln L + 0.016069(3) + 0.011(1)\frac{\ln L}{L} + \mathrm{O}\left(\frac{1}{L}, \frac{(\ln L)^2}{L^2}\right). \qquad (3.11)$$

The absence of two terms is noteworthy here. If we write down the Callan-Symanzik equation for $\Gamma'$ and insert eq. (2.11) and the perturbative $\beta$-function, it is trivial to infer that $m_2^a(L)$ should have no $(\ln L)^2$ divergence, as we indeed find. In fact, we observe the leading $2b_1 \ln L$ behavior with a relative accuracy of $10^{-3}$. To our knowledge, this is actually the first time that the 2-loop coefficient of the $\beta$-function is observed in lattice perturbation theory. The exact logarithm is then subtracted for the following analysis. The absence of the most leading artifact term $\propto (\ln L)^2/L$ is expected from tree-level improvement, and we can bound its coefficient by $\simeq 10^{-3}$. The coefficient of $1/L$ could not be determined reliably from $m_2^a(L)$ alone.

In a next step, we analyze $m_2^b(L)$, which is composed of 1-loop diagrams, and determine the leading divergence

$$c_t^{(1)} m_2^b(L) = -0.01008(5)\frac{\ln L}{L} + \mathrm{O}\left(\frac{1}{L}\right). \qquad (3.12)$$

We here used an $L$-dependent estimate for $c_t^{(1)}$ from a suitable combination of the 1-loop results for $m_1$, and we analyzed the $L$-dependence of the product as a whole. We see that the cancellation of $\ln L/L$ with the term in (3.11), which is necessary for full $O(1/L)$ improvement, is consistent with the data, with the more precise value coming from $m_2^b(L)$. Amusingly, this also constitutes another check on $c_t^{(1)}$.

We finally analyze the combination

$$m_2^a + c_t^{(1)} m_2^b + \left[c_t^{(1)}\right]^2 m_2^c - 2b_1 \ln L = 0.016069(3) - 0.0230(9)\frac{1}{L} + O\left(\frac{(\ln L)^2}{L^2}\right). \tag{3.13}$$

All terms of first order in the lattice spacing are now canceled in $m_2$ by choosing

$$c_t^{(2)} = -0.0115(5) \tag{3.14}$$

as the 2-loop improvement coefficient.

## 3.4 Two-loop relations between couplings

In this subsection we discuss the perturbative relation between our coupling $\alpha(q) = \bar{g}^2(L)/4\pi$ with $q = 1/L$ *in the continuum* and the bare coupling on the lattice[2] $\alpha_0 = g_0^2/4\pi$. This relation is expected to be free of large logarithms if $q$ is of order $a^{-1}$, where $a$ is the lattice cut-off length[3] associated with $g_0$. It is obtained by neglecting all inverse powers of $L$ and then setting $L = a$, for instance, in the formulas of the previous subsection,

$$\alpha(a^{-1}) = \alpha_0 + 2.543\,\alpha_0^2 + 9.00\,\alpha_0^3 + O(\alpha_0^4). \tag{3.15}$$

We see rather large coefficients here confirming the common experience, that $\alpha_0$ is not a good expansion parameters for physical quantities. For $\beta = 2.85$ we have $\alpha_0 \simeq 0.11$ and hence corrections of 28% and 11%. It has been heuristically argued [8], and also observed at the 1-loop level, that a better coupling is given by

$$\tilde{\alpha}_0 = \alpha_0/P, \tag{3.16}$$

---

[2] These results were also summarized in [9].
[3] Only for this subsection do we reintroduce $a$.

where $P$ is the mean plaquette at the corresponding bare coupling ($\tilde{\alpha}_0 \simeq 0.16$ at $\beta = 2.85$). With its expansion [11],[6]

$$P = 1 - \frac{3}{16}g_0^2 - 0.00946141 g_0^4 + \mathrm{O}(g_0^6), \qquad (3.17)$$

(3.15) resums to

$$\alpha(a^{-1}) = \tilde{\alpha}_0 + 0.1866\,\tilde{\alpha}_0^2 + 1.08\,\tilde{\alpha}_0^3 + \mathrm{O}(\tilde{\alpha}_0^4). \qquad (3.18)$$

The 1- and 2-loop corrections are about 3% each. In view of this, the smallness of the 1-loop term looks somewhat accidental.

We realize that the choice of exactly $1/a$ as the scale for $\alpha$ instead of some other number of the same order, like for instance $2\pi/a$ suggested by momentum discretization, is arbitrary. This clearly limits the significance of the size of the 1-loop term. With the 2-loop term at hand we have the possibility to fix this freedom in a more compelling way. Choosing the scale to minimize the 2-loop coefficient we derive

$$\begin{aligned} \alpha(1.45 a^{-1}) &= \tilde{\alpha}_0 - 0.2460 \tilde{\alpha}_0^2 + 0.891(1) \tilde{\alpha}_0^3 + \mathrm{O}(\tilde{\alpha}_0^4), & (3.19) \\ \alpha(10.91 a^{-1}) &= \alpha_0 - 0.2460 \alpha_0^2 + 1.226(1) \alpha_0^3 + \mathrm{O}(\alpha_0^4). & (3.20) \end{aligned}$$

The size of the coefficients is similar in both cases here. In fact, due to the larger value of $\tilde{\alpha}_0$, this scheme has bigger corrections at $\beta = 2.85$. The identical 1-loop coefficients at the minimizing scale are no coincidence. They are given analytically as $-2\pi b_1/b_0$. Another possibility is to choose a scale where the 1-loop correction vanishes. This results in the qualitatively similar relations

$$\begin{aligned} \alpha(1.17 a^{-1}) &= \tilde{\alpha}_0 + 0.951(1) \tilde{\alpha}_0^3 + \mathrm{O}(\tilde{\alpha}_0^4), & (3.21) \\ \alpha(8.83 a^{-1}) &= \alpha_0 + 1.287(1) \alpha_0^3 + \mathrm{O}(\alpha_0^4). & (3.22) \end{aligned}$$

In the light of these 2-loop results there arises the possibility that the main difference between the standard bare coupling and the modified bare coupling $\tilde{\alpha}_0$ consists of the scale that it is best used for. This point has been further clarified by comparing with our new high precision data in ref. [7].

The relation between $\alpha$ and $\alpha_{\overline{MS}}$ is discussed in [6]. The extension of the present calculation to SU(3) is planned.

# 4 Expanded gauge fixed action

In the remaining sections we report details of our computation. It will be of particular interest to those readers embarking on a similar lattice perturbative calculation. We shall refer from now on to section and equation numbers of ref. [3].

The perturbative expansion on the lattice requires to fix the gauge. This amounts to supplementing the pure gauge action $S$ in (2.4) by a gauge fixing term $S_{\text{gf}}$ together with a Fadeev-Popov ghost term $S_{\text{FP}}$ involving additional fermionic fields. We shall now discuss these contributions as series in $g_0$ up to $\mathrm{O}(g_0^2)$.

## 4.1 Some notation

We specialize the discussion in [3] to SU(2). In its Lie algebra we use the basis $T^\sigma, \sigma = +, -, 0$,

$$T^\pm = \frac{1}{2i}(\tau_1 \pm i\tau_2) \tag{4.1}$$

$$T^0 = \frac{1}{2i}\tau_3 \tag{4.2}$$

with the Pauli matrices $\tau_k$. For vector potentials $q_\mu(x) = q_{\mu,\sigma}(x)T^\sigma$ and $r_\mu(x) = r_{\mu,\sigma}(x)T^\sigma$ (sum over $\sigma$) the scalar product is then given as

$$(q,r) = \sum_{x,\mu}\{2\,q_{\mu,+}(x)\,r_{\mu,-}(x) + 2\,q_{\mu,-}(x)\,r_{\mu,+}(x) + q_{\mu,0}(x)\,r_{\mu,0}(x)\}, \tag{4.3}$$

and there is a corresponding formula for scalar fields without the $\mu$-sum. We define the raising of indices $\sigma$ by $q_\mu^\pm = 2q_{\mu,\mp}$, $q_\mu^0 = q_{\mu,0}$ and may then write

$$(q,r) = \sum_{x,\mu,\sigma} q_\mu^\sigma(x)\,r_{\mu,\sigma}(x). \tag{4.4}$$

## 4.2 Fields and boundary conditions

The abelian background field $V$ in (2.7) is parameterized by

$$V(x,k) = \mathrm{e}^{-\beta(x_0)T^0} \tag{4.5}$$

with
$$\beta(x_0) = -(2/L^2)[\eta L + (\pi - 2\eta)x_0]. \tag{4.6}$$

Its constant field strength is
$$V_{\mu\nu} = e^{\gamma_{\mu\nu}T^0} \tag{4.7}$$

with
$$\gamma_{\mu\nu} = -\gamma_{\nu\mu} = \begin{cases} (2/L^2)(\pi - 2\eta) & \text{for } \mu = 0,\ \nu = 1,2,3 \\ 0 & \text{else} \end{cases}. \tag{4.8}$$

It is convenient to parameterize the fluctuations of the lattice gauge field around $V$ by
$$U(x,\mu) = \exp\{g_0 q_\mu(x)\} V(x,\mu) \tag{4.9}$$

with $q_\mu$ in the Lie Algebra of SU(2). Our gauge fixing will lead to *all* modes of $q_\mu$ being quadratically damped. Hence in perturbation theory we only deal with infinitesimal $g_0 q_\mu$.

In the Schrödinger functional we have to integrate over $q_{0,\sigma}(x)$ for $x_0 = 0\ldots L-1$ and over $q_{k,\sigma}(x)$ for $x_0 = 1\ldots L-1$ ($k = 1,2,3$). As discussed in detail in [3], sect. 6, the notation is considerably simplified, if we define additional fields by boundary conditions. Since all fields are periodic in space, we can discuss them in terms of the Fourier transforms $\tilde{q}_{\mu,\sigma}(x_0, \mathbf{p})$,

$$\tilde{q}_{k,\sigma}(0, \mathbf{p}) = \tilde{q}_{k,\sigma}(L, \mathbf{p}) = 0 \tag{4.10}$$
$$\tilde{q}_{0,\sigma}(L, \mathbf{p}) = \tilde{q}_{0,\sigma}(L-1, \mathbf{p}) \tag{4.11}$$
$$\tilde{q}_{0,\sigma}(-1, \mathbf{p}) = \tilde{q}_{0,\sigma}(0, \mathbf{p}) \quad \text{if} \quad (\sigma, \mathbf{p}) \neq (0, 0) \tag{4.12}$$
$$\tilde{q}_{0,0}(-1, 0) = 0. \tag{4.13}$$

The Haar measure in the $q_\mu$ parameterization results in a local contribution to the action $S_\mathrm{m}$,

$$\prod_{x,\mu} dU(x,\mu) = \prod_{x,\mu} dq_{\mu,\sigma}(x) \exp\{-S_\mathrm{m}\} \tag{4.14}$$

with
$$S_\mathrm{m} = \frac{g_0^2}{12} \sum_{x,\mu,\sigma} q_\mu^\sigma q_{\mu,\sigma} + O(g_0^4). \tag{4.15}$$

The Fadeev-Popov determinant is represented with Lie algebra valued ghost fields $c(x)$ and $\bar{c}(x)$ with anticommuting coefficients. Their domain on the lattice is dictated by the gauge freedom that has to be fixed. Here $c(x), \bar{c}(x)$ vanish at $x_0 = L$ and are spatially constant at $x_0 = 0$. It is again convenient to extend $c$,

$$\tilde{c}_\sigma(-1, \mathbf{p}) = \tilde{c}_\sigma(0, \mathbf{p}) = 0 \quad \text{if} \quad (\sigma, \mathbf{p}) \neq (0, 0) \tag{4.16}$$
$$\tilde{c}_0(-1, 0) = \tilde{c}_0(0, 0), \tag{4.17}$$

and similarly for $\bar{c}$.

Even after gauge fixing the fluctuations, we still have covariance under gauge transformations of the background field. Hence the covariant difference, like

$$D_\mu c(x) = V(x, \mu) c(x + \hat{\mu}) V(x, \mu)^{-1} - c(x) \tag{4.18}$$

in the case of $c$, is a natural construct to use in the following. In components this corresponds to

$$D_\mu c_\sigma(x) = e^{i\sigma\beta(x_0)} c_\sigma(x + \hat{\mu}) - c_\sigma(x), \tag{4.19}$$

and

$$D_\mu^* c_\sigma(x) = c_\sigma(x) - e^{-i\sigma\beta(x_0)} c_\sigma(x - \hat{\mu}) \tag{4.20}$$

is the corresponding covariant backward difference.

## 4.3 Gluonic action

We consider now the action (2.4) as a function of $q$ and expand

$$S(U) = S(V) + S^{(0)}(q) + g_0 S^{(1)}(q) + g_0^2 S^{(2)}(q) + \mathrm{O}(g_0^3). \tag{4.21}$$

Note that the classical part $S(V)$ starts at $\mathrm{O}(g_0^{-2})$, but also contains higher order terms due to improvement,

$$S(V) = \left\{ g_0^{-2} + \frac{2}{L} [c_t^{(1)} + g_0^2 c_t^{(2)} + \ldots] \right\} \Gamma_0. \tag{4.22}$$

The simple factor $2/L$ follows from the constancy of the action density of $V$.

The part of the total action that is gaussian in $q$ is given as

$$S^{(0)} + S_{\mathrm{gf}} = (q, \Delta_1 q), \tag{4.23}$$

where the positive linear operator $\Delta_1$ is defined in [3], sect. 6.4, and will be discussed in connection with the propagators, sect. 5 of this paper. The vertices of $\mathrm{O}(g_0, g_0^2)$ will now be given as a sum of terms in the form

$$S^{(1)} = S^{(1,a)} + S^{(1,b)} + \ldots,$$

and analogously for other contributions.

A particularly simple class of terms are those which only involve the neutral gluon components with $\sigma = 0$. They are immediately generated from

$$S(U)|_{q_{\mu,\pm}=0} = \frac{4}{g_0^2} \sum_{x,\mu,\nu} w_{\mu\nu}(x) \sin^2[(\gamma_{\mu\nu} + g_0 f_{\mu\nu})/4] \qquad (4.24)$$

with the field strength $f_{\mu\nu}(x)$,

$$f_{\mu\nu}(x) = \partial_\mu q_{\nu,0} - \partial_\nu q_{\mu,0}, \qquad (4.25)$$

where $\partial_\mu$ is the forward difference operator. The relevant parts can be arranged to give the following vertices with one to four neutral gluons:

$$S^{(1,a)} = 2c_t^{(1)} \sum_{\mathbf{x},k} \sin(\gamma_{0k}/2)\{q_{k,0}(x)|_{x_0=1} - q_{k,0}(x)|_{x_0=L-1}\}, \qquad (4.26)$$

$$S^{(2,a)} = \frac{1}{2} c_t^{(1)} \sum_{\mathbf{x},k} \cos(\gamma_{0k}/2)\{f_{0k}^2|_{x_0=0} + f_{0k}^2|_{x_0=L-1}\}, \qquad (4.27)$$

$$S^{(1,b)} = -\frac{1}{12} \sum_{x,\mu<\nu} \sin(\gamma_{\mu\nu}/2) f_{\mu\nu}^3, \qquad (4.28)$$

$$S^{(2,b)} = -\frac{1}{96} \sum_{x,\mu<\nu} \cos(\gamma_{\mu\nu}/2) f_{\mu\nu}^4. \qquad (4.29)$$

The contributions with $q_{\mu,\pm}$ coming from the plaquettes are numerous and more complicated to write down. We generated them on the computer (in Fortran) in a way to later allow the fast and safe contraction with the propagators. We here only outline the general structure and quote no lists of coefficients. Our example will now be the $(q_{\mu,-} q_{\nu,+} q_{\lambda,0})$ vertex. For a given plaquette $x\,\mu\nu$ we insert into (2.4)

$$\mathrm{tr}[U_{\mu\nu}] = \mathrm{tr}\left[e^{\gamma_{\mu\nu} T^0} e^{-g_0(1+D_\nu)q_\mu} e^{-g_0 q_\nu} e^{g_0 q_\mu} e^{g_0(1+D_\mu)q_\nu}\right] \qquad (4.30)$$

and expand in the components $q_{\mu,\sigma}$. The subset of terms of the desired structure is collected in our example in the form

$$S^{(1,c)} = \sum_{x,\mu<\nu} \sum_r V_r^{(1,c)} q_{\mu_r^-,-}(x+e_r^-)\, q_{\mu_r^+,+}(x+e_r^+)\, q_{\mu_r^0,0}(x+e_r^0). \qquad (4.31)$$

The information resides in tables[4] for the directions $\mu_r^\sigma \in \{\mu,\nu\}$ and offsets $e_r^\sigma \in \{0,\hat{\mu},\hat{\nu}\}$ and in the (complex) coeffients $V_r^{(1,c)}$. Each pair of offset and direction is coded together into an integer which is taken as input by the propagator routines, when the vertex forms part of a diagram. Apart from the $V_r^{(1,c)}$ we shall also need their $\eta$-derivatives $V_r'^{(1,c)}$, which are hence stored in another array. The sum over $r$ enumerates the nonzero contributions and has 40 terms here for plaquettes involving time and 24 terms for purely spatial ones without background field. Depending on the embedding of the vertex, the number of terms can sometimes be reduced further by the over-all reality of the result.

Other vertices contributing to $S^{(2)}$ are constructed in the same fashion,

$$S^{(2,c)} = c_t^{(1)} \sum_{x_0=0,L-1} \sum_{\mathbf{x},k} \sum_r V_r^{(2,c)} q_{\mu_r^-,-}(x+e_r^-)\, q_{\mu_r^+,+}(x+e_r^+), \qquad (4.32)$$

$$S^{(2,d)} = \sum_{x,\mu<\nu} \sum_r V_r^{(2,d)} q_{\mu_r^-,-}(x+e_r^-)\, q_{\mu_r^+,+}(x+e_r^+)\, q_{\mu_r^0,0}(x+e_r^0)\, q_{\nu_r^0,0}(x+f_r^0), \qquad (4.33)$$

and

$$S^{(2,e)} = \sum_{x,\mu<\nu} \sum_r V_r^{(2,e)} q_{\mu_r^-,-}(x+e_r^-)\, q_{\mu_r^+,+}(x+e_r^+)\, q_{\nu_r^-,-}(x+f_r^-)\, q_{\nu_r^+,+}(x+f_r^+). \qquad (4.34)$$

Each vertex has new tables although we partially use the same symbols here to outline the structure. It should be noted, that for terms in $S^{(2)}$ the programming efficiency is uncritical, since in $O(g_0^2)$ they will only be self-contracted. As we evaluate graphs in position space, the space sum is trivial due to translation invariance and the CPU effort for these contributions is negligible.

We finally collect the term from $S_\mathrm{m}$ here,

$$S^{(2,f)} = \frac{1}{12} \sum_{x,\mu} \{4\, q_{\mu,+} q_{\mu,-} + (q_{\mu,0})^2\}. \qquad (4.35)$$

---

[4] To avoid too many indices we suppress here the $\mu\nu$- and $x_0$-dependence of these tables.

## 4.4 Ghost field action

The Fadeev-Popov part of the action reads

$$S_{\text{FP}} = -(\bar{c}, d^*\delta_c q), \tag{4.36}$$

and $\delta_c q$ is the gauge variation of $q_\mu$ to first order in $c$ acting as gauge parameter. $d^*$ is the covariant divergence operator, c.f. (6.13) in [3]. The expansion consists of the gaussian part

$$S_{\text{FP}}^{(0)} = (\bar{c}, \Delta_0 c) \tag{4.37}$$

with another positive linear operator $\Delta_0$ and interaction vertices

$$S_{\text{FP}}^{(1)} = -\sum_{x,\mu} \epsilon^{\sigma\tau\rho} q_{\mu,\sigma} \left(1 + \frac{1}{2}D_\mu\right) c_\tau \, D_\mu \bar{c}_\rho \tag{4.38}$$

and

$$S_{\text{FP}}^{(2)} = \frac{1}{12} \sum_{x,\mu} \left\{ q_\mu^\sigma q_{\mu,\sigma} \, D_\mu c^\tau D_\mu \bar{c}_\tau - q_\mu^\sigma q_\mu^\tau \, D_\mu c_\sigma \, D_\mu \bar{c}_\tau \right\}. \tag{4.39}$$

All indices $\sigma, \tau, \rho$ are summed over $\{0, +, -\}$, and $\epsilon^{\sigma\tau\rho}$ is the totally antisymmetric symbol with $\epsilon^{0-+} = 2i$.

# 5 Computation of propagators in the background field

In this section we discuss the definition and numerical construction of the propagators. This is a nontrivial problem due to the presence of a finite background field.

## 5.1 Definition of the ghost propagator

The ghost propagator has nonvanishing components

$$G_\sigma(x,y) = \langle c_\sigma(x)\bar{c}^\sigma(y)\rangle_0 \quad (\text{no } \sigma\text{-sum}), \tag{5.1}$$

where $\langle .\rangle_0$ is the gaussian expectation value with the order zero action only. $G_\sigma$ is inverse to $\Delta_0$, which, with the boundary conditions quoted in the last section, is just the negative covariant Laplacian. This means, that

$$-\sum_\mu D_\mu^* D_\mu F_\sigma(x) = f_\sigma(x) \tag{5.2}$$

holds if

$$F_\sigma(x) = \sum_y G_\sigma(x,y) f_\sigma(y) \tag{5.3}$$

for any $f$, which fulfills the boundary conditions. $G_\sigma(x,y)$ has to be such that this is also the case for $F_\sigma(x)$.

The symmetries of the problem suggest a spatial Fourier ansatz

$$G_\sigma(x,y) = \frac{1}{L^3} \sum_{\mathbf{p}} e^{i\mathbf{p}\cdot(\mathbf{x}-\mathbf{y})} \tilde{G}_\sigma(x^0, y^0; \mathbf{p}), \tag{5.4}$$

where the $\mathbf{p}$-sum runs over $L^3$ values suitable for the toroidal space-lattice. The components $\tilde{G}_\sigma$ are found by solving for each $\mathbf{p}$ the one-dimensional difference equations

$$\left\{-\partial_t^* \partial_t + 6 - 2\sum_{k=1}^3 \cos[p_k + \sigma\beta(t)]\right\} \tilde{G}_\sigma(t, t'; \mathbf{p}) = \delta_{tt'}, \tag{5.5}$$

For all $(\sigma, \mathbf{p}) \neq (0, 0)$ there are equations for $t, t' = 1, \ldots L-1$ consistent with the boundary conditions. This completely determines these contributions.

The remaining component $(\sigma, \mathbf{p}) = (0, 0)$ is special, as it also contributes to the propagator and fulfills (5.5) at $t, t' = 0$, and it is fixed at this end by the Neumann condition (4.17). The explicit solution for this case is simply

$$\tilde{G}_0(t, t'; 0) = L - \max(t, t'). \tag{5.6}$$

To compute the running coupling we need the $\eta$-derivatives of all contributing quantities. Taking the $\eta$ derivative of both sides in (5.5) we derive

$$\left\{-\partial_t^* \partial_t + 6 - 2\sum_{k=1}^3 \cos[p_k + \sigma\beta(t)]\right\} \tilde{G}_\sigma'(t, t'; \mathbf{p}) =$$
$$-2\sigma\beta'(t) \sum_{k=1}^3 \sin[p_k + \sigma\beta(t)] \tilde{G}_\sigma(t, t'; \mathbf{p}). \tag{5.7}$$

This equation defines $\tilde{G}_\sigma'$ in the same way as $\tilde{G}_\sigma$ as soon as the latter is known to furnish the right hand side instead of the $\delta$-source in (5.5). Note that $\tilde{G}_0'$ vanishes.

## 5.2 Definition of the gluon propagator

The gluon propagator

$$H_{\mu\nu,\sigma}(x,y) = \langle q_{\mu,\sigma}(x) q_\nu{}^\sigma(y) \rangle_0 \quad \text{(no } \sigma\text{-sum)}, \tag{5.8}$$

is inverse to $\Delta_1$ in (6.24), (6.25) of [3]. We directly pass to Fourier space,

$$H_{\mu\nu,\sigma}(x,y) = \phi_{\mu,\sigma}(x^0) \phi^*_{\nu,\sigma}(y^0) \frac{1}{L^3} \sum_{\mathbf{p}} e^{i\mathbf{p}\cdot(\mathbf{x}-\mathbf{y}+\frac{1}{2}\bar{\mu}-\frac{1}{2}\bar{\nu})} \tilde{H}_{\mu\nu,\sigma}(x^0, y^0; \mathbf{p}), \tag{5.9}$$

with

$$\bar{\mu} = \begin{cases} 0 & \text{for } \mu = 0 \\ \hat{\mu} & \text{for } \mu = 1,2,3 \end{cases} \tag{5.10}$$

and

$$\phi_{\mu,\sigma}(t) = \begin{cases} -i & \text{for } \mu = 0 \\ e^{\frac{i}{2}\sigma\beta(t)} & \text{for } \mu = 1,2,3. \end{cases} \tag{5.11}$$

The extra phases and natural space locations at the centers of links have been chosen such that $\tilde{H}_{\mu\nu,\sigma}(x^0, y^0; \mathbf{p})$ is real due to CP invariance (see sect. 6.3).

The Fourier components $\tilde{H}_{\mu\nu,\sigma}(t, t'; \mathbf{p})$ are constructed as the solutions $Q_{\mu\nu}(t, t')$ (with fixed $\mathbf{p}$, $\sigma$) of a matrix analog of (5.5),

$$\{\mathcal{A}_{\mu\kappa}(t) Q_{\kappa\nu}(t+1, t') + \mathcal{B}_{\mu\kappa}(t) Q_{\kappa\nu}(t, t') + \mathcal{A}_{\kappa\mu}(t-1) Q_{\kappa\nu}(t-1, t')\} = \delta_{\mu\nu} \delta_{tt'}. \tag{5.12}$$

Here and in the following summations over doubly occurring spin indices are implied. The coefficient matrices $\mathcal{A}, \mathcal{B}$ are given in [3] by (D.4)–(D.10) for $\sigma = 0$ and by (D.11)–(D.17) for $\sigma = +$. They represent an invertible (for $\lambda_0 > 0$) [3] real symmetric difference operator as their components are real and $\mathcal{B}$ is a symmetric matrix. Note, that the coefficients in both cases depend on $\eta$ as well as on momentum $\mathbf{p}$ and also on the gauge parameter $\lambda_0$ contained in $S_{\text{gf}}$. The time coordinate $t(t')$ associated with $\mu(\nu)$ in (5.12) runs over $0\ldots L-1$ for $\mu(\nu)=0$ and over $1\ldots L-1$ for the other components. The component $(\mu, \sigma, \mathbf{p}) = (0, 0, 0)$ is special in this case with the explicit solution

$$\tilde{H}_{0\nu,0}(t, t'; 0) = \delta_{0\nu} (1 + \min(t, t')) \lambda_0^{-1}. \tag{5.13}$$

For the $\eta$-derivative $\tilde{H}'_{\mu\nu,\sigma}$ one again derives an equation of the same type with a source term on the right hand side involving the previously constructed $\tilde{H}_{\mu\nu,\sigma}$.

## 5.3 Numerical construction of propagators

We construct and store propagators for one pair of time coordinates $(t, t')$ at a time. In a first step this is done in (spatial) momentum space. Since none of the three space directions is distinguished, there is symmetry under permuting them, which corresponds to certain discrete rotations and reflections. Hence one gains a factor of 6 (asymptotically on large lattices) by solving (5.5) and (5.12) only for the reduced set of momenta $\mathbf{p}_1 \leq \mathbf{p}_2 \leq \mathbf{p}_3$. Inspection of the propagator equations shows that for $\sigma = 0$ there is also covariance under individual reflections of any space component, which saves another factor of 8 in these sectors. This is not a symmetry of the full theory with boundary conditions, however.

The technique of solving the difference equations for given momentum $\mathbf{p}$ will now be explained for (5.12). By two-step recursion forward and backward in time we construct two solutions $\psi_{\mu\nu}^{f,b}(t)$ of the corresponding homogeneous equation[5]. They start from

$$\psi_{0\nu}^{f}(-1) = \psi_{0\nu}^{f}(0) = \delta_{0\nu} \tag{5.14}$$
$$\psi_{k\nu}^{f}(0) = 0, \ \psi_{k\nu}^{f}(1) = \delta_{k\nu} \tag{5.15}$$

and

$$\psi_{0\nu}^{b}(L) = \psi_{0\nu}^{b}(L-1) = \delta_{0\nu} \tag{5.16}$$
$$\psi_{k\nu}^{b}(L) = 0, \ \psi_{k\nu}^{b}(L-1) = \delta_{k\nu}. \tag{5.17}$$

Eq.(5.12) expresses the fact that $Q$ is the right-inverse of a certain difference operator given by $\mathcal{A}, \mathcal{B}$. Then $Q$ is also left-inverse, corresponding to the equation

$$\{Q_{\mu\kappa}(t, t'+1)\mathcal{A}_{\nu\kappa}(t') + Q_{\mu\kappa}(t, t')\mathcal{B}_{\kappa\nu}(t') + Q_{\mu\kappa}(t, t'-1)\mathcal{A}_{\kappa\nu}(t'-1)\} = \delta_{\mu\nu}\delta_{tt'}, \tag{5.18}$$

whose homogeneous solutions are also given by $\psi^{f,b}$. If we impose now (5.12) and (5.18) for $t \neq t'$ and enforce the symmetry

$$Q_{\mu\nu}(t, t') = Q_{\nu\mu}(t', t), \tag{5.19}$$

---

[5] We exclude here the case $(\sigma, \mathbf{p}) = (0, 0)$ which has different boundary conditions and can be solved in closed form.

we conclude

$$Q_{\mu\nu}(t,t') = \begin{cases} \psi^f_{\mu\kappa}(t)\,\mathcal{W}^{-1}_{\lambda\kappa}\,\psi^b_{\nu\lambda}(t') & \text{for } t \leq t' \\ \psi^b_{\mu\kappa}(t)\,\mathcal{W}^{-1}_{\kappa\lambda}\,\psi^f_{\nu\lambda}(t') & \text{for } t \geq t' \end{cases} \qquad (5.20)$$

with $\mathcal{W}_{\kappa\lambda}$ independent of $t$ and $t'$. It is a consequence of our general arguments that there must exit a choice for $\mathcal{W}$ that makes the definition (5.20) consistent and solves (5.12), (5.18) for all $t = t'$ as well. To actually determine $\mathcal{W}$, these requirements can be shown to imply

$$\mathcal{W}_{\mu\nu} = \psi^f_{\kappa\mu}(t)\,\mathcal{A}_{\kappa\lambda}(t)\,\psi^b_{\lambda\nu}(t+1) - \psi^f_{\kappa\mu}(t+1)\,\mathcal{A}_{\lambda\kappa}(t)\,\psi^b_{\lambda\nu}(t). \qquad (5.21)$$

This is a Wronskian form that is easily seen to be $t$-independent on the basis of (5.12). In the derivation one needs that $\psi^f_{\kappa\mu}(t)\,\mathcal{A}_{\kappa\lambda}(t)\,\psi^f_{\lambda\nu}(t+1)$ is symmetric in $\mu,\nu$. This follows from its antisymmetric part being another Wronskian, which vanishes at the boundary due the boundary conditions. Also $\mathcal{W}$ is simplest when computed for $t = 0$ which gives after some algebra (using (5.12) and $\mathcal{A}_{j0} = 0$)

$$\mathcal{W}_{0\nu} = \mathcal{A}_{0\kappa}(0)\psi^b_{\kappa\nu}(1) + (\mathcal{B}_{00}(0) + \mathcal{A}_{00}(-1))\psi^b_{0\nu}(0) \qquad (5.22)$$
$$\mathcal{W}_{k\nu} = -\mathcal{A}_{jk}(0)\psi^b_{j\nu}(0). \qquad (5.23)$$

The computation of the ghost propagators proceeds along the same lines, but it is much simpler due to the absence of the spin matrix structure.

It will be seen later, that for the most time consuming diagrams a summation in position space will be highly advantageous (in fact mandatory to get to L=32). Hence, after computing and storing the momentum space propagators, we Fourier-transform them to position space after constructing all **p**-components by use of the symmetries. In position space, we then found the defining equations to hold essentially with machine precision. To transform we did not use fast Fourier transform, but the ordinary one. Therefore this step (for the whole set of $(t,t')$-values) is of computational complexity $L^6$ (instead of $L^5 \log(L)$ with FFT). Although the final summations are of order $L^5$ at most, the propagators still consume only a small part of the CPU-time for $L \leq 32$. The advantage of not using FFT is the much simpler programming as we cannot restrict ourselves to $L$ which are powers of two, but rather want all L (including the odd ones) for the final analysis of the $L$-dependence. In position space we found it more convenient to use the permutation symmetry differently from before. For the gluon propagator we produce for all values of $\mathbf{x} - \mathbf{y}$ the components $\mu\nu = 00, 01, 10, 11, 12$ only. The gain factor 8 for $\sigma = 0$ from reflections is maintained through the transformation by writing special Fourier routines for functions that have definite parity in all arguments.

An initial attempt to construct propagators by (over-)relaxation directly in position space was abandoned. When trying to make use of the symmetries by identifying symmetry-related components in the iteration, we found it hard to still control efficiency and convergence.

# 6 Summation of relevant contributions

We now write down all contributions to $\Gamma$ that are of $O(g_0^2)$ in terms of the vertices to be contracted with the propagators defined in the preceding sections. As a practical matter, we discuss issues of speed and how symmetries may be used to reduce the number of terms in the sum.

## 6.1 Contributions to $\Gamma_2$

By combining now the expansion of

$$S_{\text{total}} = S + S_{\text{gf}} + S_{\text{FP}} + S_{\text{m}} \tag{6.1}$$

with (2.10) and (2.11) we have

$$\Gamma_2 = \frac{2}{L} c_t^{(2)} \Gamma_0 + \left\langle S_{\text{total}}^{(2)} - \frac{1}{2} \left( S_{\text{total}}^{(1)} \right)^2 \right\rangle_0. \tag{6.2}$$

If we now exhibit the improvement structure analogously to (3.2),

$$\Gamma_2 = \Gamma_2^a + c_t^{(1)} \Gamma_2^b + \left[ c_t^{(1)} \right]^2 \Gamma_2^c + c_t^{(2)} \Gamma_2^d, \tag{6.3}$$

the contributions arise as follows:

$$\begin{aligned}
\Gamma_2^a &= \left\langle S^{(2,b)} + S^{(2,d)} + S^{(2,e)} + S_{\text{m}}^{(2)} + S_{\text{FP}}^{(2)} \right. \\
&\quad \left. - \frac{1}{2} \left\{ S^{(1,b)} + S^{(1,c)} + S_{\text{FP}}^{(1)} \right\}^2 \right\rangle_0, \tag{6.4} \\
c_t^{(1)} \Gamma_2^b &= \left\langle S^{(2,a)} + S^{(2,c)} - S^{(1,a)} \left( S^{(1,b)} + S^{(1,c)} + S_{\text{FP}}^{(1)} \right) \right\rangle_0, \tag{6.5}
\end{aligned}$$

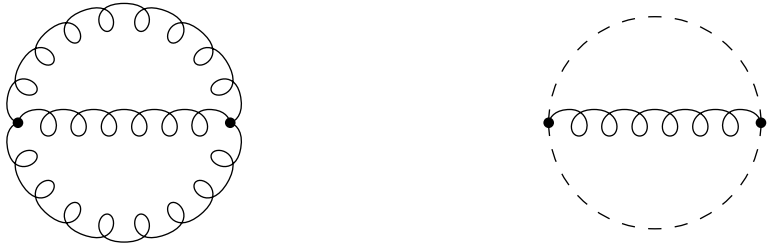

Figure 1: Expensive graphs ("Big Mac"). The dashed line represents a ghost propagator, the other ones gluonic fluctuations.

$$\left[c_t^{(1)}\right]^2 \Gamma_2^c = -\frac{1}{2}\left\langle \left(S^{(1,a)}\right)^2 \right\rangle_0, \quad (6.6)$$

$$\Gamma_2^d = \frac{2}{L}\Gamma_0. \quad (6.7)$$

Only spatial momentum zero contributes to $\Gamma_2^c$, and the relevant component of the propagator is

$$\tilde{H}_{kl,0}(t,t',0) = \delta_{kl}[\min(t,t') - tt'/L]/\cos(\gamma_{0k}/2) \quad (6.8)$$

and leads to the closed expression

$$\Gamma_2^c = -12\sin(\gamma_{0k}/2)\tan(\gamma_{0k}/2)L^2(L-2). \quad (6.9)$$

## 6.2 Efficiency and accuracy

The computational effort to compute $\Gamma_2$ is by far dominated by the contributions to $\Gamma_2^a$ from iterated 3-point vertices that are shown in graphical form in fig. 1. Each of the two vertices contains a sum over the lattice. Because of spatial translation invariance, the total double sum has $O(L^5)$ terms when carried out in position space, with a prefactor of $O(100^2)$ from the number of terms per site in each vertex. In momentum space, the complexity would be $L^8$. A term of the structure shown in fig. 2 is $O(L^7)$ in momentum space and $O(L)$ in position space which is hence clearly favoured.

As we sum a very large number of terms we have to care about roundoff errors. An obvious precaution, which also leads to a more readable code, is to introduce a hierarchy of accumulators. Apart from this we performed the sum for all but the largest lattices in 32 bit accuracy in addition to the quoted 64 bit results. The $L$-dependence of the discrepancy has a recognizable smooth trend superimposed with fluctuations, and an envelope can be estimated. We then expect the 64 bit results to have about 8 more significant decimal places than 32 bits.

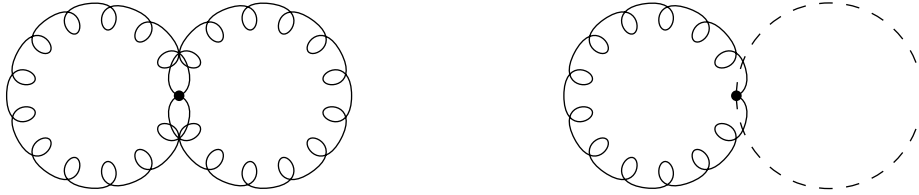

Figure 2: Simple second order graphs.

## 6.3  CP and T invariance

A possibility to still gain in efficiency is the use of further symmetries to reduce the number of summands. We implemented this after first verifying the symmetry relations in a full summation on smaller systems. The use of invariance under permutations of the three spatial directions for the construction of propagators was already discussed in sect. 5.

A factor two reduction is possible by the the use of CP invariance. Action (2.4) and background field (2.7) are invariant under

$$U(x,0) \rightarrow U^*(x',0) \tag{6.10}$$
$$U(x,k) \rightarrow U^{*-1}(x'-\hat{k},k) \tag{6.11}$$

with $x' = (x_0, -\mathbf{x})$ for $x = (x_0, \mathbf{x})$, which induces[6]

$$q_{0,\sigma}(x) \rightarrow -q_{0,-\sigma}(x') \tag{6.12}$$
$$q_{k,\sigma}(x) \rightarrow \mathrm{e}^{i\sigma\beta(x_0)} q_{k,-\sigma}(x'-\hat{k}). \tag{6.13}$$

---

[6]Periodicity is to be used to move $\mathbf{x}$ back into the range of labels chosen, $x_k = 0 \ldots L-1$, for instance.

and
$$c_\sigma(x) \to -c_{-\sigma}(x'). \tag{6.14}$$

The last relation follows from the fact that $c$ can be viewed as a gauge parameter. It makes $S_{\text{FP}}$ CP-invariant.

The actual reduction due to symmetries is now explained for the example of the left ("big mac") diagram of fig. 1. Each perturbative term in the position space form can be associated with its geometric origin, plaquettes in this case. The sum with contractions as shown can therefore be written schematically as

$$\text{diagram} = \sum_{\mu<\nu,t,\mathbf{x}} \sum_{\mu'<\nu',t',\mathbf{x}'} f(\mu,\nu,t,\mathbf{x}|\mu',\nu',t',\mathbf{x}'). \tag{6.15}$$

One sees the two sets of arguments referring to the vertices. In the process of symmetry reduction, the various terms get different multiplicities which have to be incorporated in a special array and will not be mentioned in each case. An example is the symmetry of the diagram, which allows to restrict to $t \leq t'$ if the non-diagonal terms are multiplied by two. Next, spatial translation invariance allows to pin the second plaquette to $\mathbf{x}' = 0$. Given invariance under permutations of directions, we found it simpler not to reduce the $\mathbf{x}$-sum, but to replace the 36 different $\mu, \nu, \mu', \nu'$ configurations by only 8 with appropriate multiplicities. CP corresponds to the reflection of the spatial coordinate of either term, which is given by the center of the plaquette $\mathbf{x} + (\bar{\mu} + \bar{\nu})/2$ (c.f. (5.10)). In the sum with $\mathbf{x}' = 0$ this implies identical contributions from $\mathbf{x}$ and $-\mathbf{x} - \bar{\mu} - \bar{\nu} + \bar{\mu}' + \bar{\nu}'$, which saves (roughly) a factor two in the number of different terms.

The last symmetry to be discussed is time reversal, which is more delicate. We define the T-reflection as

$$U(x,0) \to U^{-1}(x' - \hat{0}, 0) \tag{6.16}$$
$$U(x,k) \to U(x',k) \tag{6.17}$$

where now $x' = (L - x_0, \mathbf{x})$. The boundaries are swapped, and the background field is not simply invariant. We observe however the change

$$\beta(L-t) = -\beta(t) - 2\pi/L. \tag{6.18}$$

If we now follow the reflection by a gauge transformation

$$V(x,\mu) \to \Omega(x) V(x,\mu) \Omega^{-1}(x + \hat{k}) \tag{6.19}$$

with
$$\Omega = \exp\left(-T^0(x_1 + x_2 + x_3)2\pi/L\right) i\tau_1, \qquad (6.20)$$

then $V$ is invariant. We call this operation $\tilde{T}$. Note that $\Omega$ is antiperiodic in space which is admissible, as gauge fields are left periodic.

This symmetry of the original theory is still violated by our precise way of gauge fixing in eq. (6.3) of [3]. As a consequence, the symmetry is not manifest and not immediately useful for the reduction of the sum. We hence replaced this condition by another possible choice, namely

$$\omega(x)|_{x_0=0} = -\omega(x)|_{x_0=L} = \kappa, \qquad (6.21)$$

with constant $\kappa$. This is immediately inherited by $c$, which now has to be constant in space with equal and opposite values at $x_0 = 0, L$.

For the gluons we derive following [3]

$$\tilde{q}_{0,0}(-1,0) = -\tilde{q}_{0,0}(L,0) = \frac{1}{2}(\tilde{q}_{0,0}(0,0) - \tilde{q}_{0,0}(L-1,0)) \qquad (6.22)$$

and

$$\tilde{H}_{0\nu,0}(t,t';0) = \delta_{0\nu}\left(\frac{L+1}{4} - \frac{1}{2}|t - t'|\right)\lambda_0^{-1}. \qquad (6.23)$$

instead of (5.13). In addition we extend $c$ to $x_0 = -1, L+1$ in such a way that $D_\mu c$ obeys the same boundary conditions as $q_\mu$. For the ghost propagator (5.6) is replaced by

$$\tilde{G}_0(t,t';0) = \frac{L}{4} - \frac{1}{2}|t - t'|. \qquad (6.24)$$

With these changes $\tilde{T}$ invariance is manifest, and terms related by reflections of $t$ and $t'$ (taking into account the natural t-location of vertices) in the above example become pairwise identical which (asymptotically) saves another factor of two.

The stability of results under this change of boundary conditions and the manifest emergence of time reversal symmetry may also be viewed as an additional check of the computation.

**Acknowledgements:** We would like to thank Burkhard Bunk, Martin Lüscher and Peter Weisz for helpful discussions. R.N. was partially supported by DOE under grant DE-FG02-90ER40542.